\documentclass[ aip,
 amsmath,amssymb,
 reprint,%
]{revtex4-1}

\usepackage{lineno,hyperref,mathtools}
\usepackage{graphicx}
\usepackage{hyperref}

\DeclareMathAlphabet{\mathpzc}{OT1}{pzc}{m}{it}

\newcommand{\al}{\alpha}
\newcommand{\om}{\omega}

\newcommand{\tk}{\widetilde{k}}
\newcommand{\tom}{\widetilde{\omega}}
\newcommand{\tal}{\widetilde{\alpha}}
\newcommand{\ox}{\overline{x}}

\newcommand{\prt}{\partial}

\newcommand{\sn}{\mathrm{sn}}

\begin{document}

\title{Evolution of wave pulses in fully nonlinear shallow-water theory}

\author{S. K. Ivanov} \affiliation{Institute of Spectroscopy,
  Russian Academy of Sciences, Troitsk, Moscow, 108840, Russia}
\affiliation{Moscow Institute of Physics and Technology, Institutsky
  lane 9, Dolgoprudny, Moscow region, 141701, Russia}
\author{A. M. Kamchatnov} \affiliation{Institute of Spectroscopy,
  Russian Academy of Sciences, Troitsk, Moscow, 108840, Russia}
\affiliation{Moscow Institute of Physics and Technology, Institutsky
  lane 9, Dolgoprudny, Moscow region, 141701, Russia}

\begin{abstract}
We consider evolution of wave pulses with formation of dispersive shock waves
in framework of fully nonlinear shallow-water equations. Situations of initial
elevations or initial dips on the water surface are treated and motion of the
dispersive shock edges is studied  within the Whitham theory of modulations.
Simple analytical formulas are obtained for asymptotic stage of evolution of
initially localized pulses. Analytical results are confirmed by exact numerical
solutions of the fully nonlinear shallow-water equations.
\end{abstract}

\maketitle

\section{Introduction}

Although the shallow-water theory is a classical subject of investigations with a huge
number of papers devoted to it, it still remains  very active field of research with
many important applications (see, e.g., books and review articles
\cite{debnath-94,johnson-97,aosl-07,ma-10,paldor-15,apm-2018}
and references therein). Even in its simplest formulation in one-dimensional
geometry, when one neglects dissipation effects and non-uniformity of the basin's
bottom, the interplay of nonlinearity and dispersion effects leads to quite complicated
wave patterns and many approximate models were suggested for their description.
For example, if the nonlinearity and dispersion effects are taken into account in
the lowest approximation and one considers a one-directional propagation of the wave,
then its dynamics is governed by the celebrated Korteweg-de Vries (KdV)
equation~\cite{kdv-1895} and one can distinguish the following typical wave patterns.
{\it (i)} Transition between two different values of water elevation is accomplished
by an undular bore \cite{bl-54} which theory in the framework of the Whitham theory of
modulations \cite{whitham-65,whitham-74} was developed by Gurevich and Pitaevskii
\cite{gp-73}. {\it (ii)} A localized pulse of positive elevation develops a transitory
dispersive shock wave (DSW) \cite{Gur91,Gur92,Kry92,wright,tian} which evolves
eventually into a sequence of bright solitons \cite{karpman-67} accompanied by some
amount of linear radiation. {\it (iii)} If the initial pulse forms a dip in the water
surface, then it cannot evolve into a sequence of bright solitons propagating along
water surface and evolves instead into a nonlinear wave packet consisting of several
regions with different characteristic features of their behavior
\cite{an-73,zm-76,as-77,Seg81,ek-93,dvz-94,ekv-01,cg-09,cg-10} (application of Whitham
theory to this problem has been recently reconsidered in Ref.~\cite{ikp-18}).
These patterns agree qualitatively with observed phenomena but
comparison with experiments (see, e.g., \cite{hs-78,clauss-99,carmo-13,tkco-16})
shows that the KdV approximation is not good enough for quantitative description
and one needs to go beyond KdV approximation.

In KdV theory, there are two small parameters which characterize the nonlinear and
dispersion effects, respectively. In typical experimental situations,
the amplitude of the wave is not very small compared with the water depth, so the corresponding
nonlinearity parameter often cannot be considered as small. Therefore considerable efforts were
directed to the derivation of the corresponding wave equations where only small dispersive
effects are assumed. One of the most popular models was first suggested and studied in
much detail by Serre~\cite{serre-53} and later the same equations were derived in
Refs.~\cite{sg-69,gn-76} what has lead to different names attached to this system
in literature. We shall call them, as some other authors, the ``Serre equations''
after their first discoverer and investigator.

Serre equations have nice periodic and soliton solutions, so the Whitham theory
of modulations can be formulated in framework of the standard approach
\cite{whitham-65,whitham-74}. However, on the contrary to the KdV equation theory,
in the Serre equation case the Whitham modulation equations cannot be transformed to
the diagonal Riemann form and lack of Riemann invariants hinders applications of the
Whitham modulation theory to description of waves obeying the Serre
equations. In spite of that, some important consequences of the Whitham theory can
be derived without knowledge of the Riemann invariants. In particular, a smooth part
of the wave is typically described by two equations of the dispersionless approximation
which admits transformation to the Riemann diagonal form, and the same situation
takes place in the Serre equations theory. Hence, if a
one-directional (simple wave) propagation is considered, then one of the Riemann
invariants is constant. If this simple wave breaks with formation of the DSW,
then, as was first supposed by Gurevich and Meshcherkin \cite{gm-84}, the flows at
both edges of the DSW have the same values of the corresponding dispersionless
Riemann invariant. As Gurevich and Meshcherkin remarked, this rule ``plays the
same role in dispersive hydrodynamics as the shock Rankine-Hugoniot condition
in ordinary dissipative hydrodynamics'' and it ``does not depend on the nature
of dispersive effects''. This Gurevich-Meshcherkin conjecture and its generalizations
play an important role in the general theory of DSWs.

Specific dispersive properties of the system under consideration enter into the general
theory in the form of Whitham's ``number of waves'' conservation law \cite{whitham-65,whitham-74}
\begin{equation}\label{eq1}
  \frac{\prt k}{\prt t}+\frac{\prt \om(k)}{\prt x}=0,
\end{equation}
where $k=2\pi/L$ is the wave number, $L$ is the wavelength of the nonlinear periodic wave,
$\om=kV$, $V$ being the phase velocity of the periodic wave. This equation is valid along
the whole DSW and in the limit of vanishing wave amplitude the function $\om(k)$ becomes
a dispersion law of linear waves known from a linear theory. Naturally, besides wave
number $k$, it depends also on the parameters of the background smooth flow, and evolution
of all these parameters, including value of $k$ at the corresponding edge of the DSW, obeys
the limiting form of the Whitham modulation equations. A profound observation was made by
El~\cite{{El05}} who noticed that at the boundary with the simple wave these parameters depend
on a single parameter, hence Eq.~(\ref{eq1}) transforms to the ordinary differential
equation which can be integrated and the value of the corresponding integration constant
can be determined from the Gurevich-Meshcherkin closure condition. Thus, one can calculate
the wave number at the small-amplitude edge of the DSW what is enough for finding its
speed equal to the corresponding group velocity $v_g(k)=d\om/dk$ equal to the velocity of
propagation of the small-amplitude edge in evolution of bores emerging from initial
step-like discontinuities.

The situation at the opposite soliton edge of the DSW is more intricate. The number of
waves conservation law (\ref{eq1}) loses here its meaning since both $k$ and $\om$ vanish
at this edge. Nevertheless, one can include into consideration the old common wisdom
(see, e.g., \cite{ai-77,dkn-03}) that
a soliton propagates with the same velocity as its tails and if these tails decay
exponentially $\propto\exp[-\widetilde{k}|x-V_st|]$, then such a small-amplitude tail
obeys the same linear approximation of the evolution equations, as a linear propagating wave,
so that the exponential law can be obtained from a
harmonic linear wave $\propto\exp[i(kx-\om(k)t]$ by means of a simple substitution
\begin{equation}\label{eq2}
  V_s={\tom(\tk)}/{\tk},\qquad \tom(\tk)\equiv-i\om(i\tk),
\end{equation}
where $\tk$ has a meaning of the inverse half-width of the soliton and $V_s$ is its velocity.
However, $\tk$ and $\tom$ do not satisfy the ``solitonic counterpart''
\begin{equation}\label{eq4}
  \frac{\prt \tk}{\prt t}+\frac{\prt \tom(\tk)}{\prt x}=0
\end{equation}
of Eq.~(\ref{eq1}) even at the soliton edge of the DSW. To overcome this difficulty,
El studied the soliton limit of Eq.~(\ref{eq1}) and found that under certain assumptions
this limiting equation can also be transformed to the ordinary differential equation
which can be again solved with the boundary condition determined from the
Gurevich-Meshcherkin condition. Thus, the inverse half-width $\tk$ at the soliton
edge is found as well as the soliton velocity provided the parameters of the smooth
solution at this edge are already known. Since the parameters of the smooth flows at both
edges of the DSW are known in problems related with self-similar evolution of step-like
initial discontinuities, this El's scheme \cite{El05} works perfectly well in
considerations of this type of problems and it has found a number of applications
\cite{egkkk-07,ep-11,hoefer-14,hek-17} including the theory of undular bores in the
Serre equations theory~\cite{egs-06}. It is worth noticing that El's method can be
formulated effectively as an ``extrapolation'' of a solution of the limiting Whitham
equations from one edge of the DSW to the opposite edge where this limiting form is
not correct anymore. This simplified formulation provides some advantages in
generalizations of the El method, although one should remember that the
statements of the original formulation are strict in framework of the Whitham
method and of such additional assumptions as, e.g., the validity of the
Gurevich-Meshcherkin closure condition.

The integrals of modulation equations along the limiting characteristics of the DSW
edges hold not only for the initial step-like distributions. They remain correct in any
situations with breaking of simple waves and this observation allowed El and coauthors in
Ref.~\cite{egs-08} to develop the method for finding the number of solitons evolved
from initial conditions admitting the final state of the pulse evolution in the form of
solitons trains. The distribution function of solitons amplitudes was also found.
This method was applied in Ref.~\cite{egs-08} to investigation of asymptotic stage of
evolution of positive pulses for the Serre system. This generalizes the Karpman
formula \cite{karpman-67} and its analogues for other integrable equations (see, e.g.,
\cite{kku-02,kku-03}) to non-integrable situations. However, intermediate stages
of evolution often observed in actual experiments with formation of DSWs remained
inaccessible for analytical description.

To go beyond the initial step-like type of problem of Ref.~\cite{egs-06} or asymptotic
limit of Ref.~\cite{egs-08}, one has to
use some additional information about properties of the Whitham modulation equations
at the edges of DSWs. In particular, it is well known that the characteristic
velocities of the Whitham system reduce at the DSW edges, first, to the smooth flow velocity
at the edge and, second, either to the group velocity at the small-amplitude edge or to the
soliton velocity at the soliton edge. Besides that, we often know also the simple-wave
solution adjacent to the DSW. As was shown in Ref.~\cite{gkm-89} for the KdV equation
case, this information together with the known limiting expressions for Whitham
equation allows one to find the law of motion of the corresponding small-amplitude
edge for the general form of the simple-wave initial condition. This idea was used also
in Ref.~\cite{kamch-18} for finding the low of motion of the soliton edge of the DSW
in the theory of another integrable Gross-Pitaevskii (or NLS) equation. To extend this
method to non-integrable equations, we can resort to El's method of calculation of
the characteristic velocities and to combine it with well-known hodograph form of
the Whitham equations (see, e.g., \cite{kamch-2000}). As was shown in Ref.~\cite{kamch-19},
such an approach permits one to find some most important properties of pulse
evolution in simple enough analytic form and the analytical results were confirmed
by comparison with exact numerical solutions of such prototypical equations as
generalized KdV equation and generalized NLS equation. In this paper, we shall apply
the method of Ref.~\cite{kamch-19} to investigation of evolution of initial simple-wave
pulses in the Serre equations theory.

\section{Serre equations, their solutions and dispersionless limit}

We consider the Serre equations in standard non-dimensional form
\begin{equation} \label{eq5}
\begin{split}
    & h_t + (hu)_x = 0,   \\
    & u_t + uu_x + h_x = \frac{1}{3h} \left[  h^3 (u_{xt} + uu_{xx} - (u_x)^2) \right]_x,
\end{split}
\end{equation}
where $h$ is the total depth and $u$ is the horizontal velocity averaged over the
water layer depth. Looking for the travelling wave solution in the form $h=h(\xi)$,
$u=u(\xi)$, $(\xi=x-Vt)$, we find (see, e.g., \cite{egs-06}):
\begin{equation}\label{eq6}
  \begin{split}
  & h(\xi)=h_3-(h_3-h_2)\sn^2\left[\frac12\sqrt{\frac{3(h_3-h_1)}{h_1h_2h_3}}\,\xi,m\right],\\
  & u(\xi)=V-\frac{h_1h_2h_3}{h(\xi)},\qquad 0<h_1\leq h_2\leq h_3,
  \end{split}
\end{equation}
where
\begin{equation}\label{eq6b}
  m=\frac{h_3-h_2}{h_3-h_1}
\end{equation}
and $h(\xi)$ oscillates in the interval
\begin{equation}\label{eq7}
  h_2\leq h(\xi)\leq h_3.
\end{equation}
Thus, the periodic solution of the Serre equations is parameterized by four constant parameters
$h_1,h_2,h_3,V$ in terms of which all other characteristic variabled can be expressed.
In particular, the wavelength is given by the formula
\begin{equation}\label{eq8}
  L=4\sqrt{\frac{h_1h_2h_3}{3(h_3-h_1)}}K(m),
\end{equation}
where $K(m)$ is the elliptic integral of the first kind.

According to the Gurevich-Pitaevskii approach \cite{gp-73}, a DSW can be approximated
by a modulated periodic solution with slowly varying parameters $h_1,h_2,h_3,V$
which change little in one wavelength $L$. In typical situations, at one its edge,
where $h_3-h_2\ll h_2$, the DSW degenerates into the harmonic linear wave
\begin{equation}\nonumber
  \begin{split}
   h& \approx h_2+\frac12(h_3-h_2)\cos\left[\frac1{h_2}\sqrt{3\left(\frac{h_2}{h_1}-1\right)}(x-Vt)\right],\\
  u& \approx V-\sqrt{h_1}+\frac{\sqrt{h_1}}2\left(\frac{h_3}{h_2}-1\right)\\
  &\times\cos\left[\frac1{h_2}\sqrt{3\left(\frac{h_2}{h_1}-1\right)}(x-Vt)\right]
  \end{split}
\end{equation}
If we denote the background depth $h_2=h_0$, the background flow velocity
$V-\sqrt{h_1}\equiv\om/k-\sqrt{h_1}=u_0$, and the wavenumber of the harmonic wave
$$
k=\frac1{h_2}\sqrt{3\left(\frac{h_2}{h_1}-1\right)},
$$
then we get at once the well-known dispersion law
\begin{equation}\label{eq9}
  \om(k)=u_0k+k\sqrt{\frac{h_0}{1+h_0^2k^2/3}}
\end{equation}
obtained from the Serre equations linearized with respect to the background state
of water flow.

At the opposite edge for $h_2\to h_1$ the DSW transforms into a train of solitons
where the single soliton solution has the form
\begin{equation}\label{eq10}
  h=h_1+\frac{h_3-h_1}{\cosh^2\left[\frac1{2h_1}\sqrt{3\left(1-\frac{h_1}{h_3}\right)}(x-V_st)\right]}.
\end{equation}
This is a bright soliton propagating along uniform water depth $h_1=h_0$ and background
flow with velocity $u_0=V_s-\sqrt{h_3}$. If we introduce the inverse half-width of the soliton
\begin{equation}\label{eq12a}
\tk=\frac1{h_0}\sqrt{3\left(1-\frac{h_0}{h_3}\right)},
\end{equation}
then the soliton velocity can be represented as
\begin{equation}\label{eq11}
  V_s=\frac{\tom}{\tk},\qquad \tom=u_0\tk+\tk\sqrt{\frac{h_0}{1-h_0^2\tk^2/3}},
\end{equation}
where $\tom$ is related with the linear dispersion law (\ref{eq9}) by the identity (\ref{eq2}).
If the soliton propagates along a quiescent background ($u_0=0$), then its amplitude is related
with its velocity by the equation
\begin{equation}\label{eq11a}
  a\equiv h_3-h_1=V_s^2-h_1.
\end{equation}

In a modulated wave the parameters $h_1,h_2,h_3,V$ become slow functions of $x$ and $t$
and their evolution is governed by the Whitham modulation equations which can be obtained
by averaging the conservation laws \cite{whitham-65}.
Although Eq.~(\ref{eq4}) is not fulfilled, generally speaking, even along the
soliton edge of DSW, one can check \cite{kamch-19} that this equation is correct
for integrable KdV and NLS evolution equations for situations, when DSW propagates
into ``still'' medium. In completely integrable situations of KdV and NLS equations
type where the Whitham equations can be cast to the Riemann diagonal form, such
situations were called in Ref.~\cite{gkm-89} as ``quasi-simple waves'' with two Riemann
invariants constant. In non-integrable case the Riemann invariants of Whitham equations do
not exist anymore, but one can generalize the notion of quasi-simple wave as such a wave which
formal dispersionless representation is given by a multi-valued solution with two dispersionless
Riemann invariants constant (see Ref.~\cite{kamch-19}). Then we can assume that Eq.~(\ref{eq4})
is fulfilled along the soliton edge and in its vicinity. Along this edge
Eq.~(\ref{eq4}) reduces to the ordinary differential equation and its extrapolation to
the whole DSW yields with account of appropriate boundary condition the value of
$\tk$ at the soliton edge. For the initial step-like pulses, this procedure is
equivalent to the El method \cite{El05} and the advantage of formulated here
interpretation is that it can be applied to arbitrary simple-wave initial
pulses.

In dispersionless limit, the Serre equations reduce to the shallow water equations
\begin{equation} \label{eq13}
\begin{split}
     h_t + (hu)_x = 0,   \qquad
     u_t + uu_x + h_x = 0,
\end{split}
\end{equation}
which describe evolution of a smooth pulses outside the DSW. After introduction of
the Riemann invariants
\begin{equation}\label{eq14}
  r_+=\frac{u}2+\sqrt{h},\qquad r_-=\frac{u}2-\sqrt{h},
\end{equation}
Eqs.~(\ref{eq13}) can be written in the form
\begin{equation} \label{eq15}
\begin{split}
    & \frac{\prt r_+}{\prt t} + v_+(r_+,r_-)\frac{\prt r_+}{\prt x} = 0,    \\
    & \frac{\prt r_-}{\prt t} + v_-(r_+,r_-)\frac{\prt r_-}{\prt x} = 0,
\end{split}
\end{equation}
where
\begin{equation}\label{eq16}
\begin{split}
  & v_+(r_+,r_-)=\frac12(3r_++r_-),\\
  & v_-(r_+,r_-)=\frac12(r_++3r_-)
  \end{split}
\end{equation}

It is well known that any localized initial pulse with initial distributions
of $h(x,0)$, $u(x,0)$ different from some constant values $h_0,u_0$ on a finite
interval of $x$ evolves eventually into two separate pulses propagating in
opposite directions. For example, formation of such pulses from an initial
hump of elevation $h(x,0)$ has been recently considered in another physical
(nonlinear optics) context in Ref.~\cite{ikp-19} where the non-dispersive evolution
is also governed by the system (\ref{eq13}). These two pulses are called
simple-wave solutions in which one of the Riemann invariants $r_{\pm}$ is constant.
In this paper, we suppose that the initial state belongs to such a class of
simple waves. To be definite, we
assume that $r_-=u/2-\sqrt{h}=-\sqrt{h_0}=\mathrm{const}$, that is we
consider the right-propagating wave. If we denote $\sqrt{h}=c$,
$\sqrt{h_0}=c_0$, then we get
\begin{equation}\label{eq17}
  u=2(c-c_0),\quad r_+=2c-c_0,\quad v_+=3c-2c_0.
\end{equation}
The second equation (\ref{eq15}) is satisfied identically by virtue of
our assumption and the first one takes the form
\begin{equation}\label{eq18}
  c_t+(3c-2c_0)c_x=0
\end{equation}
with well-known solution
\begin{equation} \label{eq19}
    x - (3c - 2c_0)t = \ox(c),
\end{equation}
where $\ox(c)$ is the function inverse to the initial
distribution of the local sound velocity $c(x,0)$.

Now we can proceed to discussion of DSWs formed after wave breaking of such pulses.

\section{Positive pulse}

\begin{figure}[t] \centering
\includegraphics[width=9cm]{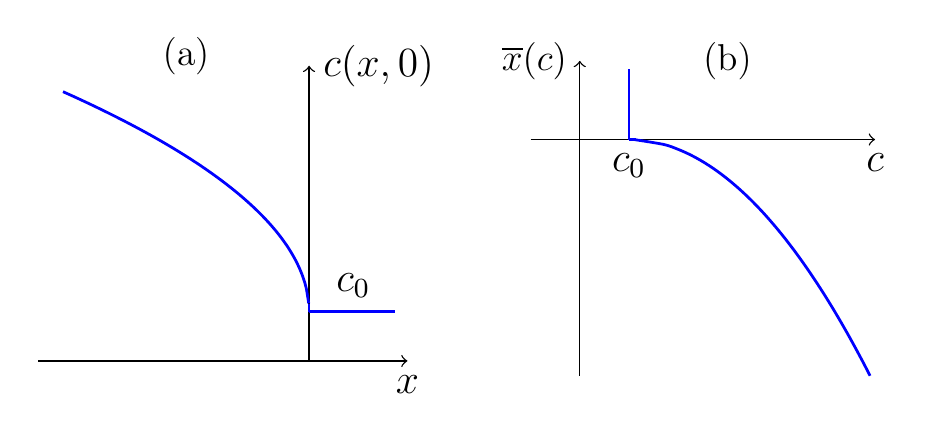}
\caption{(a) The initial distribution of the local sound velocity $c(x,0)$
in a monotonous positive pulse. (b) Inverse function $\ox(c)$.}
\label{Fig1}
\end{figure}

At first we shall consider a monotonous initial distribution
\begin{equation}\label{c_init}
c(x,0)=
\begin{cases}
c_0+\widetilde{c}(x) & \mbox{if}\quad x<0, \\
c_0 & \mbox{if}\quad x\ge 0,
\end{cases}
\end{equation}
where $\widetilde{c}(x)$ monotonously increases with decrease of $x$; see Fig.~\ref{Fig1}(a).
In Serre equations theory, the dispersion and nonlinearity lead to formation of
wave patterns qualitatively similar to those in the KdV equation theory. Hence,
we can see at once that if the initial distribution of the local sound velocity
$c$ has the form of a hump (``positive pulse''), then the wave
breaking occurs at the front of the pulse with formation of solitons at the front
edge of the DSW and with small-amplitude edge at the boundary with the smooth part
of the pulse described by Eq.~(\ref{eq19}) (rear edge of the DSW). Therefore, we
can find by the method of Ref.~\cite{kamch-19} the law of motion of the rear
small-amplitude edge of the DSW.

To this end, we assume that in vicinity of this edge Eq.~(\ref{eq1}) must be
satisfied where $k=k(c)$ is an unknown function to be found, and $u=2(c-c_0)$
(see Eq.~(\ref{eq17})). It is convenient
to represent the function $k=k(c)$ in another form with the use of the dispersion
relation (\ref{eq9}),
\begin{equation}\label{7.1}
  \om(k)=uk+kc(1+c^4k^2/3)^{-1/2}=k[u+c\al(k)],
\end{equation}
that is the function
$$
\al(k)=(1+c^4k^2/3)^{-1/2},\qquad (0<\al<1),
$$
measures deviation of the dispersion law from the linear non-dispersive law $\om=(u+c)k$,
and then we have
\begin{equation}\label{7.2}
  k(c)=\frac{\sqrt{3}}{c^2}\sqrt{\frac1{\al^2}-1},
\end{equation}
where $\al=\al(c)$ is an unknown yet function. In the simple wave solution the
Riemann invariant $r_+=r_+(c)$ and $v_+=u+c$ are functions of $c$ only, so that the
first equation (\ref{eq15}) can be cast to the form
\begin{equation}\label{7.3}
  \frac{\prt c}{\prt t}+(u+c)\frac{\prt c}{\prt x}=0.
\end{equation}
Substitution of Eqs.~(\ref{7.1})-(\ref{7.3}) into Eq.~(\ref{eq1}) yields after simple
transformations the ordinary differential equation
\begin{equation}\label{7.4}
  \frac{d\al}{dc}=-\frac{\al(4-\al)(1+\al)}{c(1+\al+\al^2)}
\end{equation}
which actually coincides with the equation obtained in Ref.~\cite{egs-06} where as an
independent variable was used $h=c^2$. Now we ``extrapolate'' this equation to the
entire DSW and find that it must be solved with the boundary condition at the
soliton edge
\begin{equation}\label{7.7}
  \al(c_0)=1,
\end{equation}
since here the distance between solitons tends to infinity, hence $k(c)\to0$ as
$c\to c_0$, and we infer (\ref{7.7}) from (\ref{7.2}). Integration of Eq.~(\ref{7.3})
with the boundary condition (\ref{7.7}) is elementary and it yields
\begin{equation}\label{7.8}
  c(\al)=\frac{c_0}{2^{1/5}3^{21/20}}\cdot\frac{(1+\al)^{1/5}(4-\al)^{21/20}}{\al^{1/4}}.
\end{equation}
The function $c(\al)$ decreases monotonously in the interval $0<\al<4$ and takes the value
$c_0$ at $\al=1$.

\begin{figure}[t] \centering
\includegraphics[width=9cm]{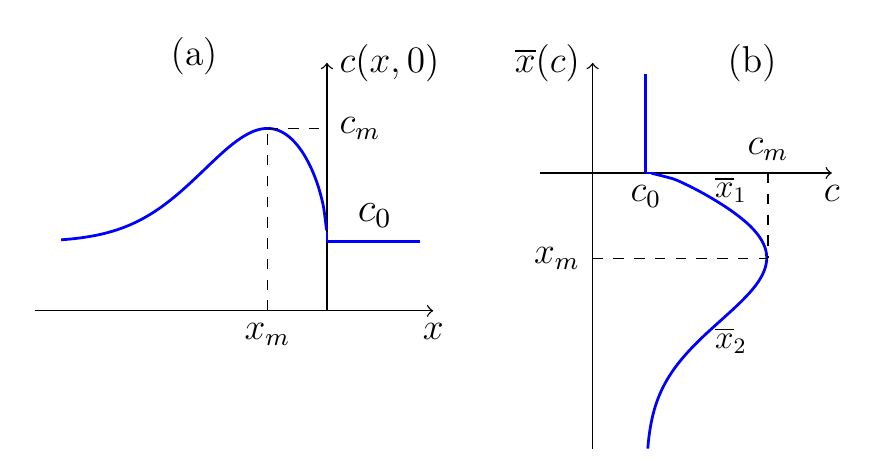}
\caption{(a) The initial distribution of the local sound velocity $c(x,0)$
in a non-monotonous pulse. (b) Inverse function $\ox(c)$ is two-valued and consist of
two branches $\ox_1(c)$ and $\ox_2(c)$.}
\label{Fig2}
\end{figure}

Now we can turn to finding the law of motion of the small-amplitude edge of the DSW which
propagates with the group velocity equal to, as one can easily obtain,
\begin{equation}\label{7.9}
  v_g=\frac{d\om}{dk}=2[c(\al)-c_0]+\al^3c(\al).
\end{equation}
According to Whitham \cite{whitham-65,whitham-74} (see also \cite{El05,kamch-19}), this is
one of the characteristic velocities of the Whitham modulation equations in the small-amplitude
limit. Another characteristic velocity must coincide, evidently, with the dispersionless
velocity $v_+=3c-2c_0$ (see Eq.~(\ref{eq18})) and the corresponding Whitham equation
separates from the other equations and yields the solution coinciding at the small-amplitude edge
with the dispersionless solution (\ref{eq19}). The characteristic velocity (\ref{7.9}) is the
common limiting value of two Whitham velocities at the small-amplitude edge of DSW, so two 
equations of the Whitham modulation system merge into one equation. Since the value of one 
more variable is fixed by the Gurevich-Meshcherkin conjecture that the value of one of the 
dispersionless Riemann invariants is transferred through the DSW, there remains in the 
small-amplitude limit only two Whitham equations for two parameters $k=k(\al)$ and $c$.
Consequently, the classical hodograph method is applicable in which $x$ and $t$ are considered 
as functions of these two parameters $k,c$, and one of these hodograph transformed equations 
must correspond to the characteristic velocity (\ref{7.9}),
\begin{equation}\label{7.10}
  \frac{\prt x}{\prt c}-[2(c-c_0)+c\al^3]\frac{\prt t}{\prt c}=0.
\end{equation}
This equation must be compatible with the solution (\ref{eq19}) of non-transformed
limiting Whitham equations, and this condition gives us the differential equation
\begin{equation}\label{7.11}
  c(1-\al^3)\frac{dt}{dc}+3t=-\frac{d\ox}{dc}
\end{equation}
for the function $t=t(c)$, where $\ox(c)$ is the function inverse to the initial distribution
of $c(x)$; see Fig.~\ref{Fig1}(b). Since we already know the relationship (\ref{7.8}) between $c$ and $\al$,
it is convenient to transform Eq.~(\ref{7.11}) to the independent variable $\al$. To simplify
the notation, we introduce the function
\begin{equation}\label{7.12}
  \Phi(\al)=\left.\frac{d\ox}{dc}\right|_{c=c(\al)}
\end{equation}
and obtain
\begin{equation}\label{7.13}
  \al(1-\al^2)(4-\al)\frac{dt}{d\al}-3t=\Phi(\al).
\end{equation}
Its solution is given by
\begin{equation}\label{7.14}
\begin{split}
  t(\al)&=\frac{\al^{3/4}(4-\al)^{1/20}}{(1-\al)^{1/2}(1+\al)^{3/10}}\\
  &\times \int_1^{\al}\frac{\Phi(z)dz}{z^{7/4}(1-z)^{1/2}(4-z)^{21/20}(1+z)^{7/10}},
  \end{split}
\end{equation}
where it is assumed that the wave breaking takes place at $t=0$ and at the front edge
of the pulse where $\al=\al(c_0)=1$. When $t=t(\al)$ is found,
the coordinate of the small-amplitude edge is given by
\begin{equation}\label{7.15}
  x_L(\al)=[3c(\al)-2c_0]t(\al)+\ox[c(\al)].
\end{equation}
The formulas (\ref{7.14}) and (\ref{7.15}) define in a parametric form the law of
motion $x=x_L(t)$ of the small-amplitude edge.

Now let us turn to a localized positive pulse when the initial distribution $c(x)$
has a single maximum $c_m$ at $x=x_m$ and $c(x)\to c_0$ fast enough as $x\to-\infty$
and $c(x)\to c_0$ with vertical tangent line at $x\to-0$ (see Fig.~\ref{Fig2}(a)).
This means that we suppose again for
convenience that the wave breaks at the moment $t=0$. Now there are two branches $\ox_{1,2}(c)$
of the inverse function with $x_m<\ox_1(c)<0$ and $-\infty<\ox_2(c)<x_m$ (see Fig.~\ref{Fig2}(b)).
The above formulas obtained for monotonous initial distribution are applicable as long as
the small-amplitude edge propagates along the branch $\ox_1(c)$, so that the solution can
be derived by replacement of the function $\Phi(\al)$ by $\Phi_1(\al)=\left. d\ox_1/dc\right|_{c=c(\al)}$.
When the small-amplitude edge reaches the maximum $c_m$ at the moment
\begin{equation}\label{7.16}
\begin{split}
  t_m&=\frac{\al_m^{3/4}(4-\al_m)^{1/20}}{(1-\al_m)^{1/2}(1+\al_m)^{3/10}}\\
  &\times \int_1^{\al_m}\frac{\Phi_1(z)dz}{z^{7/4}(1-z)^{1/2}(4-z)^{21/20}(1+z)^{7/10}},
  \end{split}
\end{equation}
where $\al_m$ is the root of the equation $c(\al_m)=c_m$, then the equation (\ref{7.13}) with
$\Phi(\al)$ replaced by $\Phi_2(\al)=\left. d\ox_2/dc\right|_{c=c(\al)}$ should be solved with
the boundary condition
\begin{equation}\label{7.17}
  t(\al_m)=t_m.
\end{equation}
Hence, for $t>t_m$ the motion of the small-amplitude is determined by the formulas
\begin{equation}\label{8.18}
\begin{split}
  &t(\al)=\frac{\al^{3/4}(4-\al)^{1/20}}{(1-\al)^{1/2}(1+\al)^{3/10}}\\
  &\times \Bigg\{\int_1^{\al_m}\frac{\Phi_1(z)dz}{z^{7/4}(1-z)^{1/2}(4-z)^{21/20}(1+z)^{7/10}}\\
  &+\int_{\al_m}^{\al}\frac{\Phi_2(z)dz}{z^{7/4}(1-z)^{1/2}(4-z)^{21/20}(1+z)^{7/10}}\Bigg\},\\
  &x_L(\al)=[3c(\al)-2c_0]t(\al)+\ox_2(c).
  \end{split}
\end{equation}

\begin{figure}[t] \centering
\includegraphics[width=7cm]{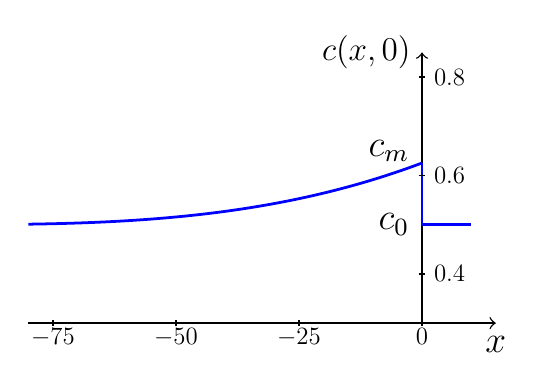}
\caption{The initial distribution of the local sound velocity $c(x,0)$
given by Eq.~(\ref{8.21}) with parameters $c_0=0.5$, $c_1=0.005$, $x_0=100$.}
\label{Fig3}
\end{figure}

\begin{figure}[t] \centering
\includegraphics[width=8cm]{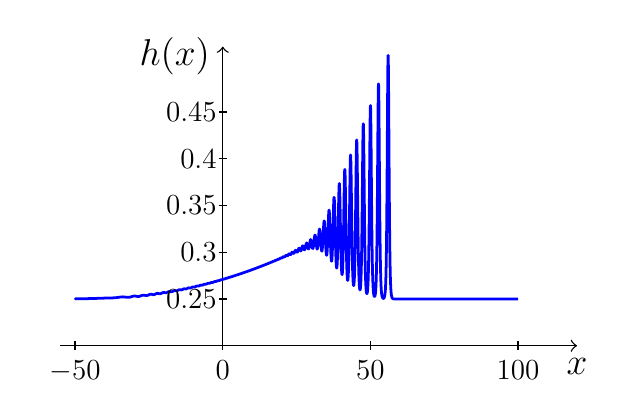}
\caption{A typical surface elevation $h(x)$ at fixed moment of time $t = 80$ in the pulse
evolved from the initial distribution (\ref{8.21}) found by numerical solution
of the Serre equations.
}
\label{Fig4}
\end{figure}

The law of motion of the soliton edge of DSW cannot be found by this method, since the
relation (\ref{eq4}) does not hold during evolution of the pulse. However, as was remarked
in Ref.~\cite{kamch-19}, this relationship can be used in vicinity of the moment when the
small-amplitude edge reaches the point with $c=c_m$. We rewrite Eq.~(\ref{eq11}) in the form
\begin{equation}\label{9.1}
  \tom(\tk)=u\tk+\tk c(1-c^4\tk^2/3)^{-1/2}=\tk[u+c\tal(\tk)],
\end{equation}
that is
$$
\tal(\tk)=(1-c^4\tk^2/3)^{-1/2},\qquad (\tal>1),
$$
and then
\begin{equation}\label{9.3}
  \tk(c)=\frac{\sqrt{3}}{c^2}\sqrt{1-\frac1{\tal^2}}.
\end{equation}
Substitution of these expressions into Eq.~(\ref{eq4}) with account of $u=u(c)$ yields the
familiar differential equation
\begin{equation}\label{9.4}
  \frac{d\tal}{dc}=-\frac{\tal(4-\tal)(1+\tal)}{c(1+\tal+\tal^2)}
\end{equation}
coinciding with Eq.~(\ref{7.4}). Now we extrapolate it to the whole DSW what allows us to
solve it with the boundary condition
\begin{equation}\label{9.5}
  \tal(c_m)=1,
\end{equation}
which means that the solitons widths become infinitely large ($\tk\to0$) and their
amplitude infinitely small at the small-amplitude edge. Then the solution of the
differential equation coincides with Eq.~(\ref{7.8}) after obvious replacements
$\al\to\tal$, $c_0\to c_m$,
\begin{equation}\label{9.6}
  c(\tal)=\frac{c_m}{2^{1/5}3^{21/20}}\cdot\frac{(1+\tal)^{1/5}(4-\tal)^{21/20}}{\tal^{1/4}}.
\end{equation}
Here $c<c_m$, $\tal$ increases with decrease of $c$ and reaches its maximal value at
$c=c_0$ which corresponds to the maximal possible value of the soliton velocity equal to
\begin{equation}\label{10.1}
  v_R=\left.\frac{\tom}{\tk}\right|_{c=c_0}=c_0\tal(c_0).
\end{equation}
This value of the leading soliton velocity in DSW is reached at large enough time when
solitons near the leading edge are well separated from each other and propagate along
the constant background with $c=c_0$. Substitution of $\tal(c_0)=v_R/c_0$ into
Eq.~(\ref{9.6}) yields the equation
\begin{equation}\label{91.2}
  \frac{v_R}{c_0}=\left(\frac{c_m}{c_0}\right)^4\left(\frac{1+v_R/c_0}{2}\right)^{4/5}
  \left(\frac{4-v_R/c_0}{3}\right)^{21/5},
\end{equation}
which determines implicitly the value of $v_R$ in terms of $c_0$ and $c_m$. This equation
coincides actually with the equation obtained in Ref.~\cite{egs-06} for velocity of the
leading soliton in a bore generated from a step-like initial distribution what should be
expected since such a distribution can be modeled by a long enough pulse with a sharp
front. When the velocity of the leading soliton is found, its amplitude is given by
(see Eq.~(\ref{eq11a}))
\begin{equation}\label{91.3}
  a=v_R^2-h_0=h_0\left[\tal^2\left(\sqrt{h_0}\right)-1\right]
\end{equation}
in agreement with the result obtained in Ref.~\cite{egs-08} for asymptotic stage of evolution
of positive pulses.

\begin{figure}[t] \centering
\includegraphics[width=6cm]{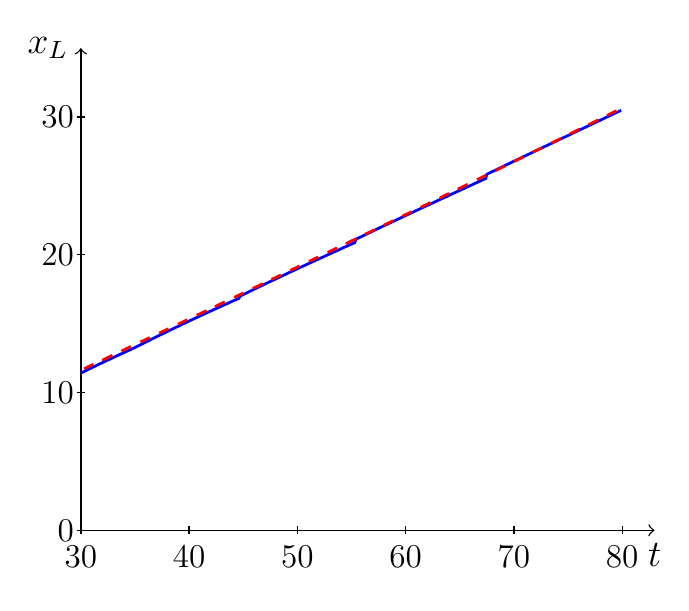}
\caption{A path of the small-amplitude edge in DSW generated by the initial distribution
(\ref{8.21}) shown in Fig.~\ref{Fig3}. The numerical results are depicted by a solid (blue) line.
The analytical curve corresponds to Eqs.~(\ref{8.18}) and it is shown by dashed (red) line.
}
\label{Fig5}
\end{figure}

It is instructive to compare the formulas of the Serre equations theory
with the results obtained in the small nonlinearity
approximation when propagation of the pulse obeys the KdV equation
\begin{equation}\label{8.19}
\begin{split}
  & u_t+\sqrt{gh_0}\left(u_x+\frac16h_0^2u_{xxx}\right)+\frac32uu_x=0,\\
  & u=2(c-c_0).
  \end{split}
\end{equation}
A similar calculation yields (see Ref.~\cite{kamch-19})
\begin{equation}\label{8.20}
  \begin{split}
  &t(c) =-\frac1{6\sqrt{c-c_0}}\Big\{\int_{c_0}^{c_m}\frac{\ox_1'(c)dc}{\sqrt{c-c_0}}
  +\int_{c_m}^{c}\frac{\ox_2'(c)dc}{\sqrt{c-c_0}}\Big\},\\
  &x_L(\al)=[3c(\al)-2c_0]t(\al)+\ox_2(c).
  \end{split}
\end{equation}
Naturally, these formulas correspond to the approximation of the function $c(\al)$
\begin{equation}\label{80.1}
\begin{split}
  & c(\al)\approx c_0[1+(1-\al)/2],\quad\text{that is}\\
  &\al(c)\approx3-2c/{c_0},
  \end{split}
\end{equation}
and can be obtained from Eqs.~(\ref{8.18}) by substitution into them of this leading
approximation of the function $\al(c)$.

In a similar way, in the KdV approximation Eq.~(\ref{91.2}) yields the asymptotic
value of the leading soliton velocity
\begin{equation}\label{80.2}
  v_R\approx 2c_m-c_0.
\end{equation}

\begin{figure}[t] \centering
\includegraphics[width=6.5cm]{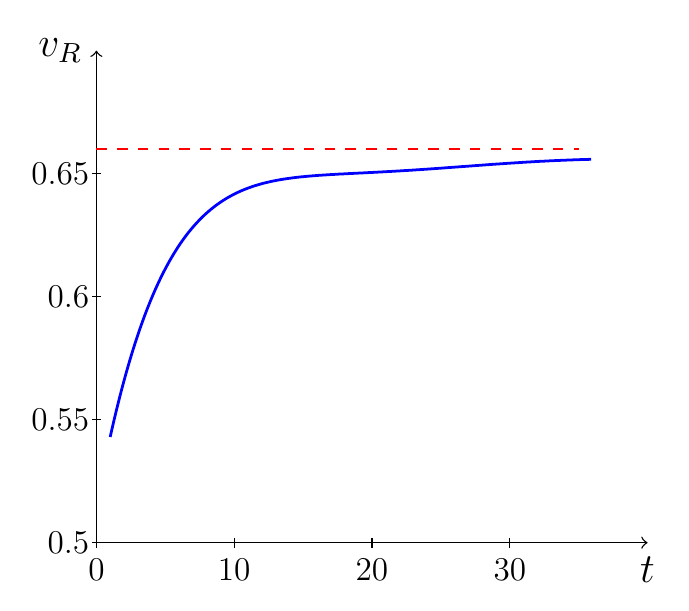}
\caption{Velocity of the soliton edge as a function of time (solid blue line).
Dashed (red) line indicates the asymptotic value of the velocity of this edge given by Eq.~(\ref{10.1}).
}
\label{Fig6}
\end{figure}

\begin{figure}[t] \centering
\includegraphics[width=8cm]{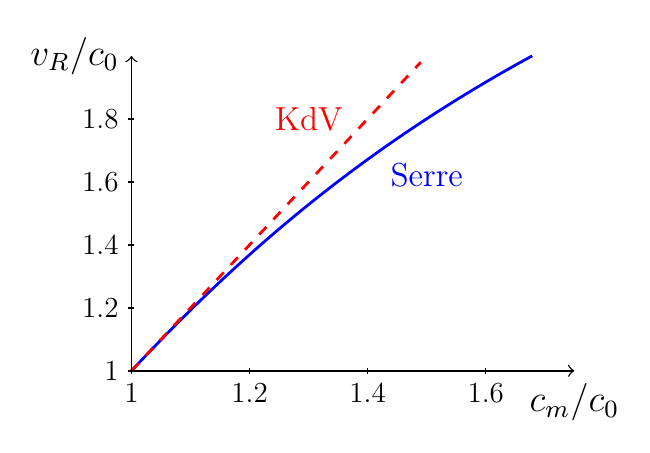}
\caption{Comparison of velocity $v_R/c_0$ of the leading soliton according to the Serre theory
(Eq.~(\ref{10.1}); solid blue line) and the KdV equation theory (Eq.~(\ref{80.2}); dashed red line)
as functions of the amplitude $c_m/c_0$ of the initial pulse.
}
\label{Fig7}
\end{figure}

We have compared our analytical approach to DSWs in the Serre equations theory with their
numerical solution for the pulse with $x_m=0$, sharp
front edge and long enough rear distribution described by the function (see Fig.~\ref{Fig3})
\begin{equation}\label{8.21}
  c(x)=c_0+[c_1(x+x_0)]^3,
\end{equation}
so that
\begin{equation}\label{8.21b}
\begin{split}
&\ox_2(c-c_0)=(c-c_0)^{1/3}/{c_1}-x_0,\\
&  \Phi_2(\al)=[c(\al)-c_0]^{-2/3}/(3c_1).
  \end{split}
\end{equation}
The typical wave pattern for the water elevation evolved from such a pulse is shown in
Fig.~\ref{Fig4}. It is worth noticing that the amplitude of the wave has the same order of
magnitude as the background depth, that is the KdV approximation is not applicable in this
case. Propagation path of the small-amplitude edge is depicted in Fig.~\ref{Fig5}.
Since the Whitham asymptotic theory does not take into account the initial stage of formation
of the DSW, the analytical curve is shifted slightly upwards to match the numerical one.
As we see, this single fitting parameter permits one to describe the whole curve very well.

Dependence of the leading soliton velocity on time is shown in Fig.~\ref{Fig6}. As we see,
it approaches with time to its asymptotic value given by Eq.~(\ref{10.1}). At last,
in Fig.~\ref{Fig7} we compare the dependence of velocity of the leading soliton on the
amplitude $c_m/c_0$ of the initial pulse according to the Serre theory (solid line) and
Korteweg-de Vries theory (dashed line). As one can see,  the difference between two
approximations increases very fast with growth of the amplitude.

Thus, our analytical approach agrees quite well with velocities of propagation of
the edges of DSW evolved from arbitrary enough initial profile of the pulse in the
Serre equation theory.

\section{Negative pulse}

\begin{figure}[t] \centering
\includegraphics[width=9cm]{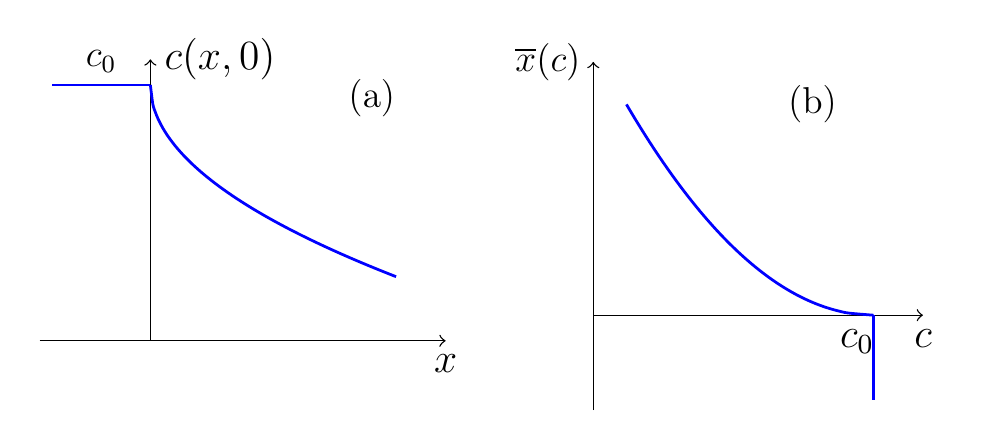}
\caption{(a) The initial distribution of the local sound velocity $c(x,0)$
in a monotonous negative pulse. (b) Inverse function $\ox(c)$.}
\label{Fig8}
\end{figure}

Here we turn to the discussion of evolution of a ``negative'' pulse with a dip in the initial
distribution of the local sound velocity $c(x)$. Again we shall start from a monotonous pulse
with the initial distribution of the local sound velocity
\begin{equation}\label{8.23}
c(x,0)=
\begin{cases}
c_0 & \mbox{if}\quad x<0, \\
c_0-\widetilde{c}(x) & \mbox{if}\quad x\ge 0,
\end{cases}
\end{equation}
which is shown in Fig.~\ref{Fig8}(a), and the inverse function $\ox(c)$ shown in Fig.~\ref{Fig8}(b).
As was indicated
in Ref.~\cite{kamch-19}, in situations of this kind Eq.~(\ref{eq4}) is fulfilled at the soliton edge
located at the boundary with the smooth part of the pulse which is described in dispersionless
approximation by Eq.~(\ref{eq19}). This smooth part acquires with time a triangle-like shape
which is well known in the KdV equation theory (see, e.g. \cite{as-77,dvz-94,ikp-18}) and in
the ``shallow water'' system derived as a dispersionless approximation in the NLS equation
theory (see, e.g., \cite{ikp-19}).
As in Refs.~\cite{El05,kamch-19}, we assume that Eq.~(\ref{eq4})
is valid also for DSWs propagating into a quiescent medium and apply this conjecture to description
of DSW evolved from the initial distribution (\ref{8.23}) according to the Serre equations. Then
Eq.~(\ref{eq4}) can be transformed in the same way as it was done in Eqs.~(\ref{9.1})-(\ref{9.4}),
but now Eq.~(\ref{9.4}) must be solved with the boundary condition
\begin{equation}\label{8.24}
  \tal(c_0)=1
\end{equation}
since in the case of negative initial pulse the small-amplitude edge of the DSW is located at the
boundary with the quiescent medium where $c=c_0$ and solitons amplitude tends to zero as $c\to c_0$.
The solution is obtained from Eq.~(\ref{9.6}) by a replacement $c_m\to c_0$, i.e.,
\begin{equation}\label{8.25}
  c(\tal)=\frac{c_0}{2^{1/5}3^{21/20}}\cdot\frac{(1+\tal)^{1/5}(4-\tal)^{21/20}}{\tal^{1/4}},
\end{equation}
where $\tal>1$. When the function $\tal(c)$ is known, the velocity of the front edge
can be represented as the soliton velocity
\begin{equation}\label{10.2}
  v_R=\frac{\tom}{\tk}=2(c-c_0)+c\tal.
\end{equation}
We notice again that $v_R$ is the characteristic velocity of the Whitham modulation equations
at the soliton edge. The corresponding Whitham equation can be written in the hodograph transformed
form
\begin{equation}\label{11.1}
  \frac{\prt x}{\prt c}-[2(c-c_0)+c\tal]\frac{\prt t}{\prt c}=0.
\end{equation}
The condition that this equation must be compatible with the solution (\ref{eq19}) of the
dispersionless approximation at the soliton edge of the DSW yields the differential equation
for the function $t=t(c)$ at this edge,
\begin{equation}\label{11.2}
  c(\tal-1)\frac{dt}{dc}-3t=\frac{d\ox}{dc}.
\end{equation}
Introducing the function $\Phi(\tal)=\left.d\ox/dc\right|_{c=c(\tal}$ we cast it to the form
\begin{equation}\label{12.1}
  \frac{\tal(\tal^2-1)(4-\tal)}{1+\tal+\tal^2}\frac{dt}{d\tal}+3t=-\Phi(\tal)
\end{equation}
with $\tal$ as an independent variable. Since we assume that the initial profile breaks at the
moment $t=0$ at the rear edge where $c=c_0$, {\it i.e.,} $\tal=1$, this equation should be solved with
the boundary condition
\begin{equation}\label{12.2}
  t(1)=0.
\end{equation}
As a result, we obtain
\begin{equation}\label{12.3}
\begin{split}
  t(\tal)=&-\frac{\tal^{3/4}(4-\tal)^{21/20}}{(\tal-1)^{3/2}(\tal+1)^{3/10}}\\
  &\times\int_1^{\tal}\frac{\Phi(z)(1+z+z^2)(z-1)^{1/2}dz}{z^{7/4}(z+1)^{7/10}(4-z)^{41/20}}.
  \end{split}
\end{equation}
The coordinate of the soliton edge is given by
\begin{equation}\label{13.1}
  x_R(\tal)=[3c(\tal)-2c_0]t(\tal)+\ox[c(\tal)],
\end{equation}
so that Eqs.~(\ref{12.3}), (\ref{13.1}) define the law of motion $x=x_R(t)$ of this edge
in parametric form.

\begin{figure}[t] \centering
\includegraphics[width=9cm]{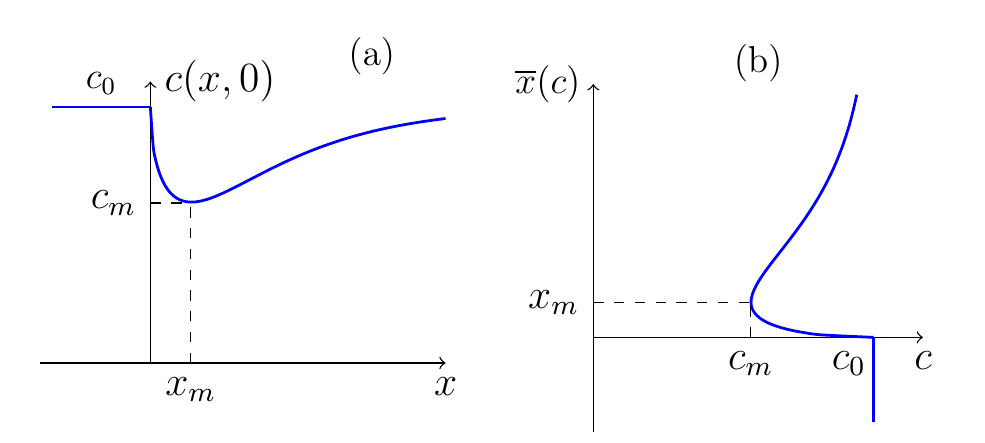}
\caption{(a) The initial distribution of the local sound velocity $c(x,0)$
in a nonmonotonous negative pulse. (b) Inverse function $\ox(c)$.}
\label{Fig9}
\end{figure}

Generalization of this theory on localized pulses (see Fig.~\ref{Fig9}(a)) is straightforward.
Now we have a two-valued inverse function which consists of two branches $\ox_1(c)$
and $\ox_2(c)$ (see Fig.~\ref{Fig9}(b)). Equations (\ref{12.3}), (\ref{13.1}) with $\Phi$
replaced by $\Phi_1(\tal)=\left.d\ox_1/dc\right|_{c=c(\tal}$ are valid up to the moment
\begin{equation}\label{13.2}
\begin{split}
  t_m=&-\frac{\tal^{3/4}(4-\tal)^{21/20}}{(\tal-1)^{3/2}(\tal+1)^{3/10}}\\
  &\times\int_1^{\tal_m}\frac{\Phi_1(z)(1+z+z^2)(z-1)^{1/2}dz}{z^{7/4}(z+1)^{7/10}(4-z)^{41/20}}
  \end{split}
\end{equation}
when the soliton edge reaches the minimum $c_m$ in the distribution of the local sound
velocity and $\tal_m$ is defined by the equation $c(\tal_m)=c_m$. After that moment $t>t_m$
the soliton edge propagates along the second branch $\ox_2(c)$ and the law of motion of
this edge is defined by the equations
\begin{equation}\label{14.1}
\begin{split}
  t(\tal)=&-\frac{\tal^{3/4}(4-\tal)^{21/20}}{(\tal-1)^{3/2}(\tal+1)^{3/10}}\\
  &\times\Bigg\{\int_1^{\tal_m}\frac{\Phi_1(z)(1+z+z^2)(z-1)^{1/2}dz}{z^{7/4}(z+1)^{7/10}(4-z)^{41/20}}\\
  &+\int_{\tal_m}^{\tal}\frac{\Phi_2(z)(1+z+z^2)(z-1)^{1/2}dz}{z^{7/4}(z+1)^{7/10}(4-z)^{41/20}}\Bigg\},\\
  x_R(\tal)&=[3c(\tal)-2c_0]t(\tal)+\ox_2[c(\tal)].
  \end{split}
\end{equation}
At asymptotically large time $t\to\infty$ we have $\tal\to1$ and
$t(\tal)\approx\mathcal{A}/(\tal-1)^{3/2}$, where
\begin{equation}\label{14.2}
  \mathcal{A}=-\frac{3^{21/20}}{2^{3/10}}
  \int\frac{\Phi(z)(1+z+z^2)(z-1)^{1/2}dz}{z^{7/4}(z+1)^{7/10}(4-z)^{41/20}},
\end{equation}
where the integral should be taken over the whole initial pulse
($\int\Phi=\int_1^{\tal_m}\Phi_1+\int_{\tal_m}^1\Phi_2$). In this limit
$c(\tal)\approx c_0[1-(\tal-1)/2]$ and simple calculation yields
\begin{equation}\label{14.3}
  x_R(t)\approx c_0t-\frac32\mathcal{A}^{2/3}t^{1/3}.
\end{equation}
This asymptotic law of motion coincides actually with the corresponding law in the KdV
equation theory but with a different value of the constant $\mathcal{A}$
(see Ref.~\cite{ikp-18}).

The method of Ref.~\cite{kamch-19} permits us to find asymptotic value of velocity of
the rear small-amplitude edge at $t\to\infty$. To this end, we have to solve
Eq.~(\ref{7.4}) with the boundary condition
\begin{equation}\label{15.1}
  \al(c_m)=1
\end{equation}
which means that at the moment when the soliton edge reaches the minimum of distribution
of $c(x)$, the wave number here given by Eq.~(\ref{7.2}) tends to zero. This gives
\begin{equation}\label{15.2}
  c(\al)=\frac{c_m}{\al^{1/4}}\left(\frac{1+\al}2\right)^{1/5}
  \left(\frac{4-\al}3\right)^{21/20}.
\end{equation}
At the small amplitude edge we have $c=c_0$ and the group velocity (\ref{7.9}) plays the role
of the left-edge velocity,
\begin{equation}\label{15.3}
  v_L=\al_0^3c_0,
\end{equation}
where $c_0=c(\al_0)$.
Substitution of $\al_0=(v_L/c_0)^{1/3}$ into Eq,~(\ref{15.2}) yields the equation
\begin{equation}\label{15.4}
  \left(\frac{v_L}{c_0}\right)^{\frac1{12}}=\frac{c_m}{c_0}\left[\frac{1+(v_L/c_0)^{\frac13}}2\right]^{\frac15}
  \left[\frac{3-(v_L/c_0)^{1/3}}3\right]^{\frac{21}{20}}
\end{equation}
which determines $v_L$ in terms of $c_0$ and $c_m$. In the KdV approximation
Eq.~(\ref{15.4}) gives
\begin{equation}\label{15.5}
  v_L\approx 6c_m-5c_0.
\end{equation}
This relation agrees with the corresponding result of Ref.~\cite{ikp-18} after its
transformation to physical variables of Eq.~(\ref{8.19}).

\begin{figure}[t] \centering
\includegraphics[width=8cm]{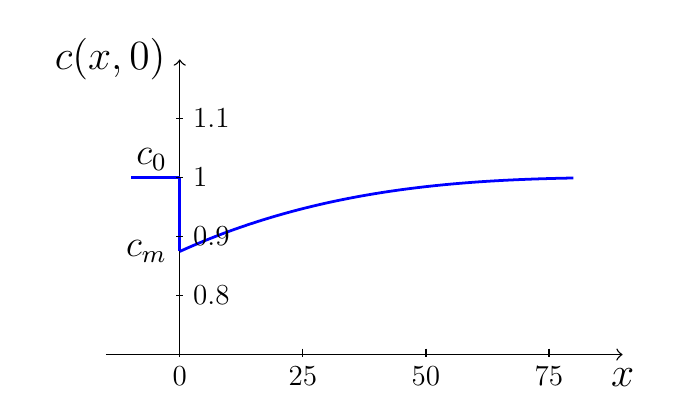}
\caption{The initial distribution of the local sound velocity $c(x,0)$
given by Eq.~(\ref{16.1}) with parameters $c_0=0.5$, $c_1=0.005$, $x_0=100$.}
\label{Fig10}
\end{figure}

\begin{figure}[t] \centering
\includegraphics[width=8cm]{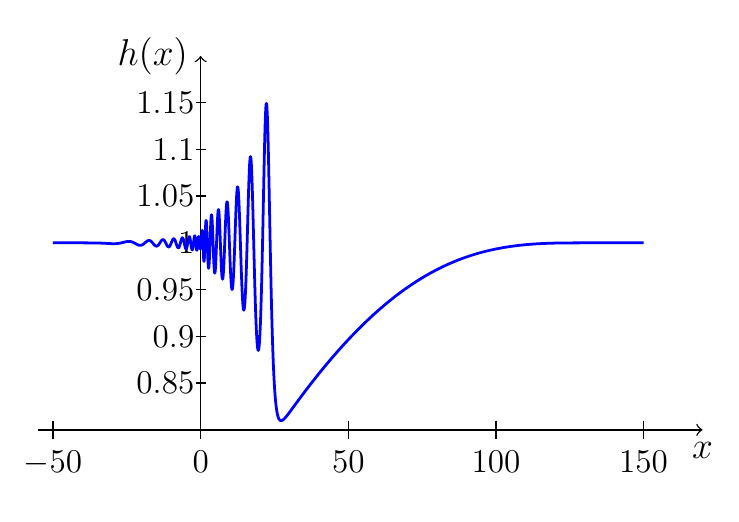}
\caption{A typical surface elevation $h(x)$ at fixed moment of time $t = 30$ in the pulse
evolved from the initial distribution (\ref{16.1}) found by numerical solution
of the Serre equations.
}
\label{Fig11}
\end{figure}

We compared our analytical approach to DSWs in the Serre equations theory with their
numerical solution for the pulse with the initial distribution of the local sound velocity
defined by the equation (see Fig.~\ref{Fig10})
\begin{equation}\label{16.1}
  c(x)=c_0-[c_1(-x+x_0)]^3,
\end{equation}
so that
\begin{equation}\label{16.2}
\begin{split}
&\ox_2(c)=x_0-(c_0-c)^{1/3}/{c_1},\\
&  \Phi_2(\al)=[c_0-c(\al)]^{-2/3}/(3c_1).
  \end{split}
\end{equation}
The typical wave pattern for the water elevation evolved from such a pulse is shown in
Fig.~\ref{Fig11}. As was mentioned above, the smooth part of the pulse acquires a triangle-like 
shape in agreement
with typical behavior of dispersionless solutions of shallow water equations (\ref{eq13}).

\begin{figure}[t] \centering
\includegraphics[width=8cm]{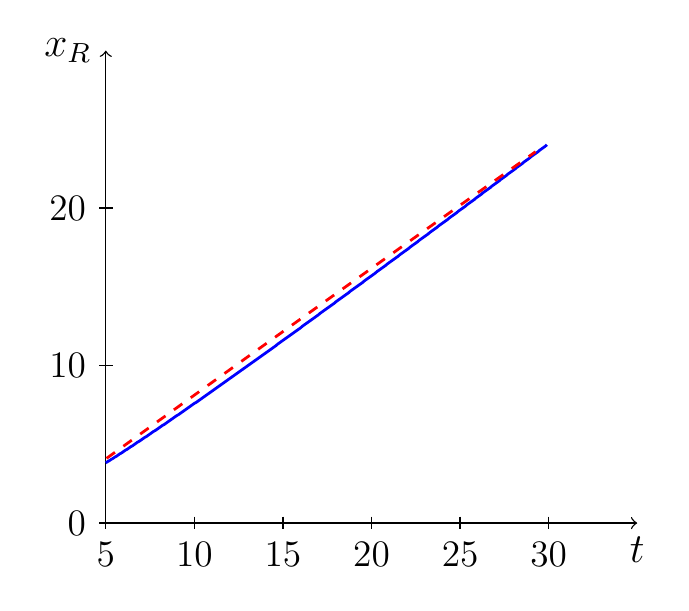}
\caption{Propagation path of the soliton edge: numerical results are shown by a solid (blue)
line and analytical calculation by a dashed (ref) line.
}
\label{Fig12}
\end{figure}

Propagation path of the soliton edge is depicted in Fig.~\ref{Fig12}.
Again the analytical curve is shifted slightly upwards to match the numerical one.
As we see, this single fitting parameter permits one to describe the whole curve very well.

\begin{figure}[t] \centering
\includegraphics[width=8cm]{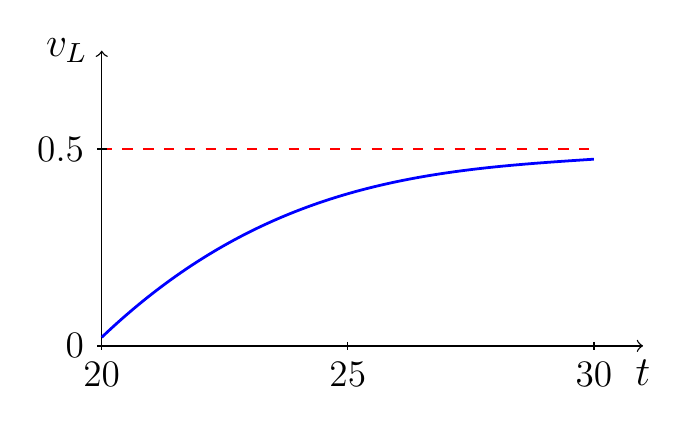}
\caption{Dependence of the velocity of the small-amplitude edge on time: numerical results
are shown by a solid (blue)
line and the analytical asymptotic value (\ref{15.3}) by a dashed (red) line.
}
\label{Fig13}
\end{figure}

\begin{figure}[t] \centering
\includegraphics[width=8cm]{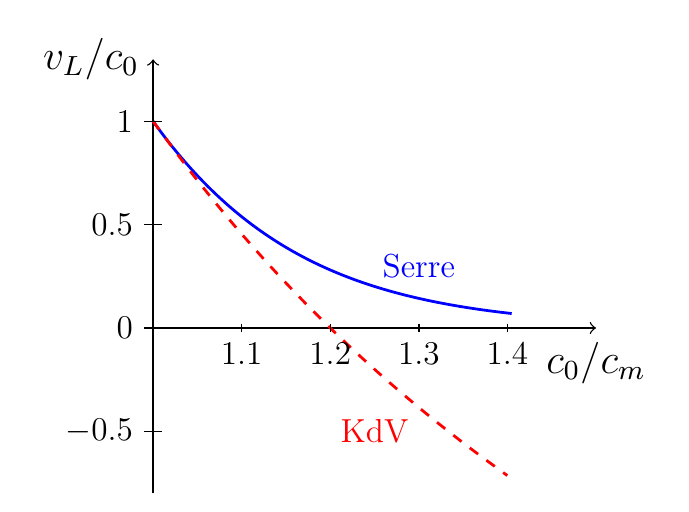}
\caption{Dependence of the velocity of the small-amplitude edge according to the Serre
theory (upper blue line) and to the KdV approximation (lower red line).
}
\label{Fig14}
\end{figure}

Dependence of the velocity of the small-amplitude edge on time is shown in Fig.~\ref{Fig13}.
As we see, it approaches with time to the asymptotic law of motion with the constant velocity
given by Eq.~(\ref{15.3}). At last,
in Fig.~\ref{Fig14} we compare the dependence of velocity of the small-amplitude edge on the
parameter $c_0/c_m$ according to the Serre theory (upper blue line) and
Korteweg-de Vries theory (lower red line). As one can see, again the difference between two
approximations increases very fast with growth of depth of the dip in the initial distribution.

\section{Conclusion}

We have studied evolution of dispersive shock waves formed after wave breaking
of simple-wave pulses in the fully nonlinear shallow-water approximation
in framework of the Serre equations model with the use of the Whitham modulation
theory. Formation of undular bores from initial step-like distributions
was studied earlier in Ref.~\cite{egs-06} and asymptotic stage of positive initial
pulses at the large enough time, when it evolves to a train of well separated solitons,
was studied in Ref.~\cite{egs-08}. The present study essentially extends
such an approach to much wider class of initial conditions with both types
of possible polarities (humps and dips in initial elevations) and to any value
of time evolution provided the Whitham method is applicable. In particular,
we have obtained the laws of motion of the small-amplitude edge of DSW evolved
from a positive pulse and the law of motion of the soliton edge of DSW evolved
from negative pulse, when velocities of these edges depend on the initial profiles
of the pulses and they are given by simple analytical formulas. Especially
useful in practical applications may be formulas for velocities of the edges
at asymptotically large time. In addition to the leading soliton velocity
generated by a positive pulse which was already found in Ref.~\cite{egs-08},
we have found the small-amplitude velocity for the same initial polarity
and expressions for both edges in the case of negative polarity.
Remarkably enough, these asymptotical velocities
depend on a single parameter given by either a certain integral over the initial
distribution, or its extremal value within the initial pulse. We hope that
the developed here theory can find applications to description of experimental
investigations of water waves of the type performed in Refs.~\cite{hs-78,clauss-99,tkco-16}.
The developed theory demonstrates quite clearly that the method suggested in
Ref.~\cite{kamch-19} is convenient enough for discussion of concrete situations
and it can be effectively applied to various models of water wave physics
and other nonlinear physics problems.

\begin{acknowledgments}
We are grateful to G.~A.~El, M.~Isoard and N.~Pavloff for useful discussions.
\end{acknowledgments}

\end{document}